\documentclass[aps,prb,twocolumn,epsfig,floatfix,showpacs,superscriptaddress,amsmath,amssymb]{revtex4}
\usepackage{amssymb}
\usepackage{amsmath}
\usepackage{latexsym}
\usepackage{graphicx}
\usepackage{dcolumn}
\usepackage{bm}
\usepackage{natbib}
\DeclareGraphicsExtensions{.eps}

\newcommand{\be}{\begin{equation}}
\newcommand{\ee}{\end{equation}}
\newcommand{\ba}{\begin{eqnarray}}
\newcommand{\ea}{\end{eqnarray}}
\newcommand{\baa}{\begin{eqnarray*}}
\newcommand{\eaa}{\end{eqnarray*}}

\begin{document}

\title{Interfacial Effects of Al-Termination on Spin Transport in Magnetic Tunnel Junctions}
\author{T. Tzen Ong}
\affiliation{Department of Applied Physics, Stanford University, Stanford CA 94305, USA}
\affiliation{IBM Research Division, Almaden Research Center, San Jose, CA 95120, USA}
\author{A. M. Black-Schaffer}
\affiliation{Department of Applied Physics, Stanford University, Stanford CA 94305, USA}
\author{W. Shen}
\affiliation{Department of Physics, Georgetown University, Washington DC 20057, USA}
\affiliation{IBM Research Division, Almaden Research Center, San Jose, CA 95120, USA}
\author{B. A. Jones}
\affiliation{IBM Research Division, Almaden Research Center, San Jose, CA 95120, USA}
\date{\today}

\begin{abstract}
Experiments have shown that the tunneling current in a Co/Al$_2$O$_3$ magnetic tunneling junction (MTJ) is positively spin polarized, opposite to what is intuitively expected from standard tunneling theory which gives the spin polarization as exclusively dependent on the density of states (DOS) at $E_F$ of the Co layers. Here we report theoretical results that give a positive tunneling spin polarization and  tunneling magnetoresistance (TMR) that is in good agreement with experiments. From density functional theory (DFT) calculations, an Al-rich interface MTJ with atomic-level disorder is shown to have a positively polarized DOS near the interface. We also provide an atomic model calculation which gives insights into the source of the positive polarization. A layer and spin dependent effective mass model, using values extracted from the DFT results, is then used to calculate the tunneling current, which shows positive spin polarization. Finally, we calculate the TMR from the tunneling spin polarization which shows good agreement with experiments.
\end{abstract}

\pacs{}

\maketitle

\section{Introduction}
\label{sec:intro}
The discovery of large tunneling magnetoresistance (TMR) in magnetic tunneling junctions\cite{Moodera95} (MTJs) has prompted great interest in developing these devices for applications such as magnetic random access memories (MRAMs) and magnetic sensors (for reviews, see e.g.~Refs.~[\onlinecite{Moodera99,Tsymbal03}]). 
The TMR should, in an ideal system, be closely related to the spin polarization of the two ferromagnetic layers sandwiching the insulator, and is typically given by the Julliere model \cite{Julliere75} as
\be
TMR \equiv \frac{R_{AP} - R_P}{R_{AP}} = \frac{2 P_1 P_2}{1 + P_1 P_2}
\label{eq:TMR Julliere}
\ee
Here $R_{AP}$ and $R_P$ denote the resistance for the anti-parallel and parallel alignments of the ferromagnetic layers, respectively, and $P_i$ (i = 1 and 2) are the bulk magnetic moments of the two layers. 
This model is based on the assumption that the spin polarization of the tunneling current, which determines the TMR, is an intrinsic property of the spin polarization in the ferromagnetic layers. Experimental results, however, do not often match this phenomenology. Most importantly, for the ferromagnetic 3d metals such as Co, which have a negative spin polarization at the Fermi level, the measured spin polarization across an alumina,  Al$_2$O$_3$, insulating barrier is positive. \cite{Meservey94}. Experiments have also demonstrated a strong dependence of spin transport on the detailed structural and electronic nature of the insulating layer and its interface with the ferromagnetic layers. \cite{LeClair.PRL.2001, LeClair.PRL.2002}

Density functional theory (DFT) calculations of O-terminated interfaces have shown a positive spin polarization for barrier distances beyond 10 \AA. \cite{Oleinik00, Oleynik03}. Additional DFT work have shown that additional oxygen atoms absorbed at the interface form strong Co-O bonds and give rise to a positive spin interface band. This band was found to be dominant in the tunneling process, and gives rise to a positive spin polarization of the tunneling current. \cite{BelashchenkoPRB05} Formation of an O-terminated interface with Co-O bonds is obtainable by moderate to long oxidation times of the Al$_2$O$_3$ layer. An increased magnetization of the interface Co atoms was found to coincide with the optimum TMR, using X-ray magnetic circular dichroism measurements. \cite{TellingJAP06} However, one recent experiment failed to detect any induced magnetic moment on the interface O-atoms,\cite{Bowen06} which should also be present. Hence, details of the physical nature of O-terminated MTJs remains open.

In this work we will instead focus on clean Al-terminated and disordered Al-rich interfaces which are experimentally accessible by not overoxidizing the junction. \cite{PlischAPL01} Earlier DFT results on Al-terminated junctions have shown that the magnetic moment on the Co interface atoms is somewhat reduced by charge transfer and screening effects\cite{Oleinik00} but Al-rich junctions have not yet been studied nor have the positive spin polarization been explained in any of these junctions.
We show here that increasing Al content at the interface can cause dramatic changes in the spin polarization of both the interface region and the tunneling current, even to the degree of changing the sign of the spin polarized tunneling current as measured experimentally. This explicitly demonstrates that oxygen absorption is not the only means to achieve positive spin polarization as was recently suggested.\cite{BelashchenkoPRB05}

More specifically, we use DFT to model Co/Al$_2$O$_3$/Co magnetic tunnel junctions with varying Al-rich interfaces to determine atomic structure, band structure, and local density of states (LDOS). 
The DFT results explicitly show the importance of the proximity effect on the LDOS of the interfacial layers. The LDOS near $E_F$ of the interfacial Co layer leaks into the first Al layer of the insulating barrier, and gives rise to a small but finite DOS near $E_F$, hence metallizing the Al layer, as has been seen experimentally.\cite{PlischAPL01} The Al layer also picks up a small ferromagnetic moment due to this proximity effect. Similarly, the LDOS and the spin polarization of the nearest Co layer is reduced. Thus, the spin DOS changes as we cross the interface, and the spin polarization actually becomes positive for Al-rich interfaces.

We use a position-dependent effective mass model to explicitly calculate the spin polarized tunneling current of our junctions. The change in the effective mass of the different bands between the electrode bulk-like layers and the interfacial layers result in better matching between the spin-up bands, compared to the spin-down bands. There is then less reflection at the interface, and hence a larger transmission coefficient for the spin-up bands ultimately resulting in a positive spin polarized tunneling current in the Al-rich junctions.

The paper is organized as follows. In Section \ref{sec:compdetails} we review the details specific to the DFT calculations. We also present the different interface structures we have studied. In Section \ref{sec:clean} we report the results for the clean Al-terminated interface as a baseline for the Al-rich interfaces. In Section \ref{sec:intscattering} we discuss interfacial scattering by excess Al atoms before we present DFT results for disordered, Al-rich interfaces in Section \ref{sec:disorderedint}. Finally in Section \ref{sec:spincurrent} we explicitly calculate the spin polarized tunneling current and the TMR based on the DFT LDOS and band structure data.

\section{DFT Computational Details}
\label{sec:compdetails}
\begin{figure}[bht]
\begin{center}
\includegraphics[height=10cm]{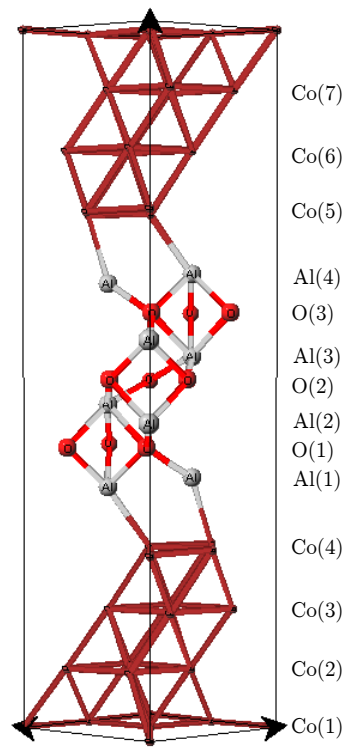}
\end{center}
\caption{Atomic structure of the clean Al-terminated interface junction. The unit vectors for the supercell as well as the layer identification tags are shown. Layers Co(1), Co(2), and Co(7) were kept fixed in order to impose bulk Co conditions.}
\label{Supercell_clean}
\end{figure}

To model the Co/Al$_2$O$_3$/Co magnetic tunnel junction we have created Co/Al$_2$O$_3$/Co supercells with various Al-terminated interfaces. Thin film Co is predominantly found in the fcc phase and experimentally it has been found that alumina grows on top of the (111) plane of fcc Co.\cite{LeClair.PRL.2002}
We have chosen a  2x2 surface unit cell of (111) fcc Co with the theoretical lattice parameter $a = 3.38$ \AA\ ($a_{exp} = 3.55$ \AA) as the base structure. In order to enforce bulk conditions in the inner cobalt layers we fix the atomic positions for the three innermost Co layers.
For the tunneling barrier we use one unit cell of $\alpha$-Al$_2$O$_3$ (corundum) with the [0001] orientation and Al-termination to model the experimentally amorphous alumina. This structure contains four Al layers with three O layers in between them. Since alumina is grown on top of cobalt we adjust the lateral dimensions of alumina to that of the cobalt. 
This is the same crystallographic orientation as used in previous DFT studies\cite{Oleinik00,Oleynik03,BelashchenkoPRB05} and gives an experimental lattice mismatch of only 6\% between the cobalt and alumina.
While most previous theoretical work have focused on oxygen rich interfaces there also exist experimental data showing Al-termination at the Co/Al$_2$O$_3$ interface.\cite{PlischAPL01} We investigate both the clean, abrupt interface with Al-termination, see Fig.~\ref{Supercell_clean}, and two different cases of Al-rich interfaces. The first case has one mixed interface layer with a 1:1 Al to Co ratio. The second case has two mixed layers with a 3:1 and 1:1 Co to Al ratio, respectively, see Fig.~\ref{Supercell_2}. We will present the results for the clean interface system in Sec.~\ref{sec:clean}, and the results for the disordered interface system with  two mixed layers in Sec.~\ref{sec:disorderedint}. We found the results for the system with one mixed layer to be an interpolation of the clean and two mixed layer interfaces.

The band structure and LDOS were calculated using the first-principle density functional theory (DFT) pseudopotential method implemented in the Vienna Ab-initio Simulation Package (VASP).\cite{Kresse96} We have used ultrasoft pseudopotentials and employed both the local density approximation (LDA) and the general gradient approximation (GGA). In general, we expect the LDA calculations to underestimate the lattice constant and the band gap  but with more consistent errors than a GGA calculation. The cut-off energy for the plane wave expansion was set to 29 Ry.
After k-point convergence tests we chose a  9x9x1 $\Gamma$-centered  k-point sampling. The atomic structure, except the fixed three innermost Co layers, was relaxed until the change in total energy between two ionic steps was less than $10^{-3}$ eV.
The LDOS per layer was calculated as a spd site-projected LDOS using the monoatomic Wigner-Seitz radii and then summed over each layer, see Figs.~\ref{Supercell_clean} and \ref{Supercell_2} for typical layout of the layers.

\section{Clean Interface}
\label{sec:clean}
Figure \ref{Supercell_clean} shows the supercell used for calculating the properties of a clean, Al-terminated interface junction. We performed both LDA and GGA calculation but found the LDA results to be more self-consistent. For reference we also used Co in the hcp structure but could not detect any significant change from the fcc structure, which was then used throughout the rest of the work.

\begin{figure}[bht]
\begin{center}
\includegraphics[width=8cm]{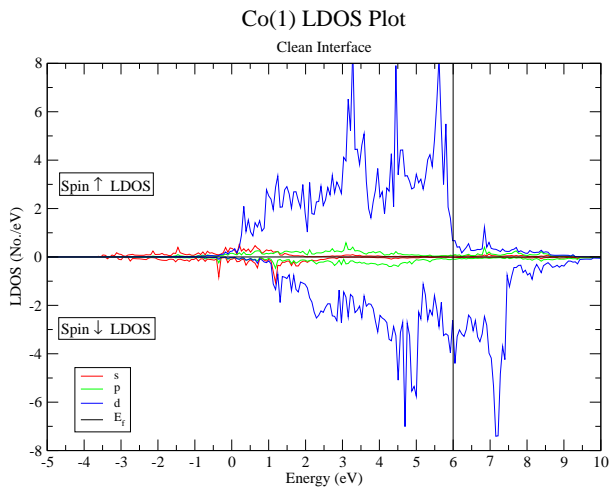}
\end{center}
\caption{LDOS for the innermost Co layer, Co(1), for the clean Al-terminated interface.  The LDOS shows typical bulk-like character as expected.}
\label{Clean_Co1_LDOS}
\end{figure}

\begin{figure}[bht]
\begin{center}
\includegraphics[width=8cm]{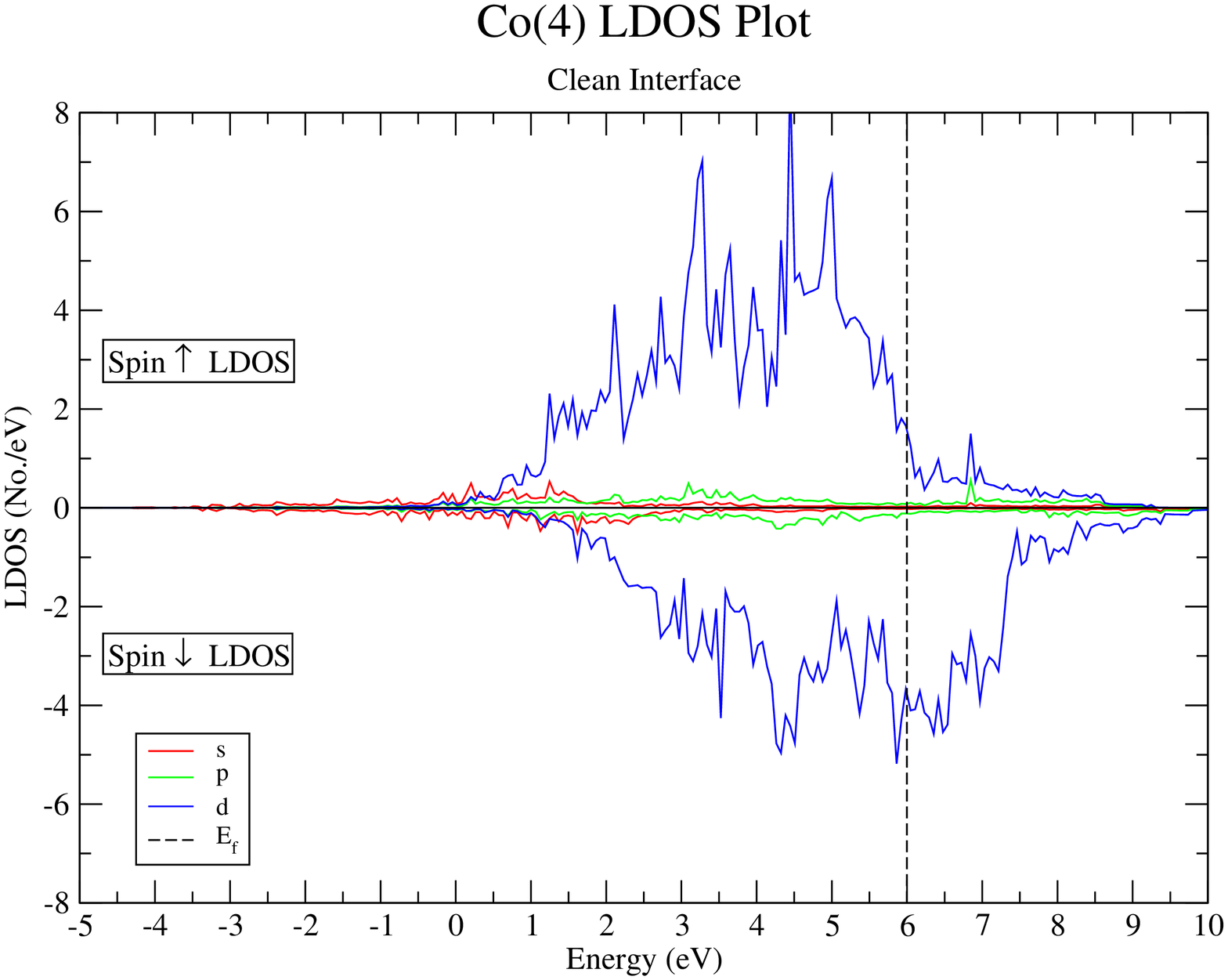}
\end{center}
\caption{LDOS for the interfacial Co layer, Co(4), for the clean Al-terminated interface.}
\label{Clean_Co4_LDOS}
\end{figure}

\begin{figure}
\begin{center}
\includegraphics[width=8cm]{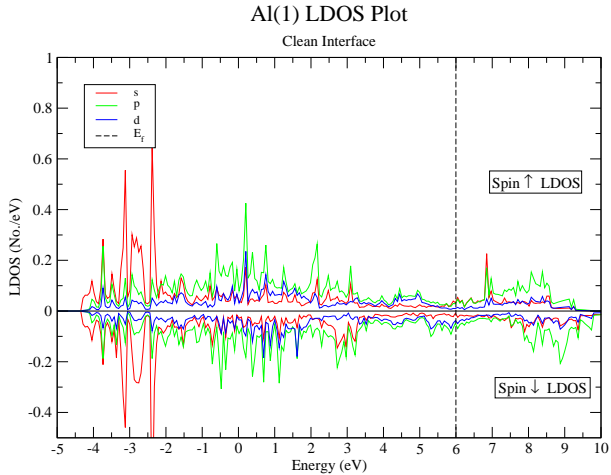}
\end{center}
\caption{LDOS for the first Al layer in the insulating barrier, Al(1), for the clean Al-terminated interface.}
\label{Clean_Al1_LDOS}
\end{figure}

Figures \ref{Clean_Co1_LDOS}-\ref{Clean_Al1_LDOS} show the LDOS for the innermost Co layer, Co(1), the interface Co, Co(4), layer, as well as the first Al layer, Al(1). As expected, the LDOS for Co(1), which is furthest from the interface, is almost identical to that of bulk Co with an exchange splitting of $\approx 1.5$ eV. The major peaks in the spin-up d-electron LDOS is below $E_F$, indicating that the spin-up d-orbitals are filled as expected. Similarly, some of the spin-down d-electron peaks are above $E_F$, showing that some of the spin-down d-orbitals are unfilled; this gives rise to the net magnetic moment. The s- and p-electrons also show some hybridization with the d-electrons, as evidenced by the slight increase in the spin-up s- and p-electron density at $\approx 3$ eV, and the spin-down s-electron and p-electron density at $\approx 4.3$ eV. 

The LDOS for Co(4), which is the layer at the interface for the clean interface, shows the effect of hybridization with the s-p electrons in the Al(1) layer next to it. The d-electrons are more delocalized and this is reflected in that the d-electron peaks are not as well-defined as in the Co(1) layer. The peak in the spin-up d band near 3 eV is reduced, and the peak near 5.6 eV has been shifted to a much smaller peak near 5 eV, and the rest of the spectral weight spectral weight has been moved to a more delocalized LDOS around 5 eV. 

There is also a slight reduction in the magnetic moment in this Co(4) layer as compared to the bulk-like Co(1). The spin polarization, $\delta_s = \frac{\rho_{\uparrow}(E_F) - \rho_{\downarrow}(E_F)}{\rho_{\uparrow}(E_F) + \rho_{\downarrow}(E_F)}$, for the Co(1) layer is -0.64, but -0.43 for the Co(4) layer. For this clean interface system, this is due to the hybridization between the Co d-electrons and the Al s-p electrons at the interface. The trend of increasing positive spin polarization as the interface is approached is seen in all three systems, i.e. in both the clean interface and the two disordered interface cases which we have modeled. The detailed behavior of the spin polarization is shown in Fig.~\ref{spin_polarization_plot}, and is discussed further in Section. \ref{sec:disorderedint}.

The Al(1) layer is the start of the insulating Al$_2$O$_3$ barrier. However, due to hybridization of the Al s-p electrons with the d-electrons in the neighbouring Co(4) layer, there is a finite DOS at $E_F$. It is only the O(1) layer next to Al(1) that shows a gap at $E_F$ and therefore the actual width of the insulating barrier is decreased by one to two atomic layers. Calculation of the current-voltage curves using both cases show that a decrease of one atomic layer gives a better fit with experimental results. \cite{Panchula_private} This means that the metallization of the Al(1) and Al(4) layers in actual samples is probably only partial, with a resulting decrease in tunneling width of a total of only one atomic layer.

\section{Interfacial Scattering and Friedel Screening Effects}
\label{sec:intscattering}
From TEM studies of MTJ structures, it can be seen that there is atomic-level disorder at the interface. \cite{Bae.Apl.2002, Portier.APL.2001} This atomic level disorder at the interface will necessarily result in scattering of the conduction electrons. Assuming Al-termination at the interface, the Al atoms will act as positively-charged impurities, because of their charge in the alumina, and this charge should get screened by the s-p and d electrons in the Co. From Friedel's sum rule, we know that the conduction electrons will screen the impurity charge, resulting in a change in the LDOS near the interface. Almost all of the screening will be due to the d-electrons as the LDOS of the d-electrons is much larger than the s- and p-electrons for Co at $E_F$. Also, since the spin-down DOS in cobalt is larger than the spin-up DOS at $E_F$, more spin-down electrons will be involved in the screening. This will result in a spin dependent change in the LDOS at the interface. Similar spin dependent charge transfer effects have also been seen in other DFT works on clean Al-terminated interfaces, where there is a larger transfer in the spin-down band than the spin-up band. \cite{Oleinik00}

We can calculate the approximate shift in $E_F$ and LDOS at $E_F$, $\delta \rho^{o}_s(E_F)$, for the spin polarized d-electron bands, and give an estimate of the spin dependent Friedel-screening effect.\cite{Doniach99} We assume one Al impurity per unit cell and use the LDOS of clean Co as shown in Fig.~\ref{Clean_Co1_LDOS}. The total phase shift for both spin-up and down bands must satisfy $\theta_{\uparrow}(E_F) + \theta_{\downarrow}(E_F) = \pi Z$, where Z is the charge of the impurity, so as to completely screen the impurity. The phase shift of each band, $\theta_s(\epsilon)$, is given by,

\begin{eqnarray}
\tan(\theta_s(\epsilon)) & = & \frac{\pi \rho_s^{o}(\epsilon)}{F_s(\epsilon) - \frac{1}{U}} \nonumber \\
F_s(\epsilon) & = & P \int \frac{\rho_s^{o}(\epsilon ')}{\epsilon - \epsilon'}{\rm   d}\epsilon'.
\end{eqnarray}

Here, U, the impurity potential, as seen by both bands is clearly the same. Since $\rho_{\downarrow}^o(E_F) > \rho_{\uparrow}^o(E_F)$, the spin-down band is expected to screen more of the impurity charge than the spin-up band. In the ``rigid-band" approximation, the shift in $E_F$ can be approximated by $\Delta_s = \tfrac{\theta_s}{\pi \rho_s^o(E_F)}$, and the change in the LDOS at $E_F$ is approximated by $\delta \rho^{o}_s(E_F) = \Delta_s \frac{\partial \rho_s^o}{\partial \epsilon}(E_F)$.  A straightforward calculation gives $\theta_{\uparrow}(E_F) = 0.42 \pi$ and $\theta_{\downarrow}(E_F) = 0.58 \pi$, which gives $\Delta_{\uparrow} = 0.30$~eV and $\Delta_{\downarrow} = 0.56$~eV, showing that the spin-down d-electrons experience a Freidel screening effect that is about twice as strong. Since the tunneling current depends strongly on the DOS at $E_F$, this change in the LDOS will have a significant effect on the spin polarized tunneling current.
 
As will be shown below, DFT results for the Al-rich interfaces clearly demonstrates the significance of hybridization and the spin dependent Friedel screening effect. In order to obtain an accurate and reliable band structures arising from this Al-induced disorder effect, we carried out DFT calculations on interfaces with an excess amount of Al incorporated at the interface. We believe the most obvious place for these excess Al atoms are to replace Co in the Co-lattice and we therefore created interfaces with Al-Co layers. As described in Section \ref{sec:compdetails} we studied both junctions with one and two Co-Al interfacial layers. The case with only one interfacial layer can qualitatively be interpreted as an interpolation between the clean and the two interfacial Co-Al layer interface so we will for clarity and compactness only report the results on the case with two interfacial Al-Co layers. The layer spin polarization in Fig.~\ref{spin_polarization_plot}, which includes the results for all three structures, clearly shows the intermediate features for the one interfacial layer structure.

\begin{figure}[bht]
\begin{center}
\includegraphics[height=10cm]{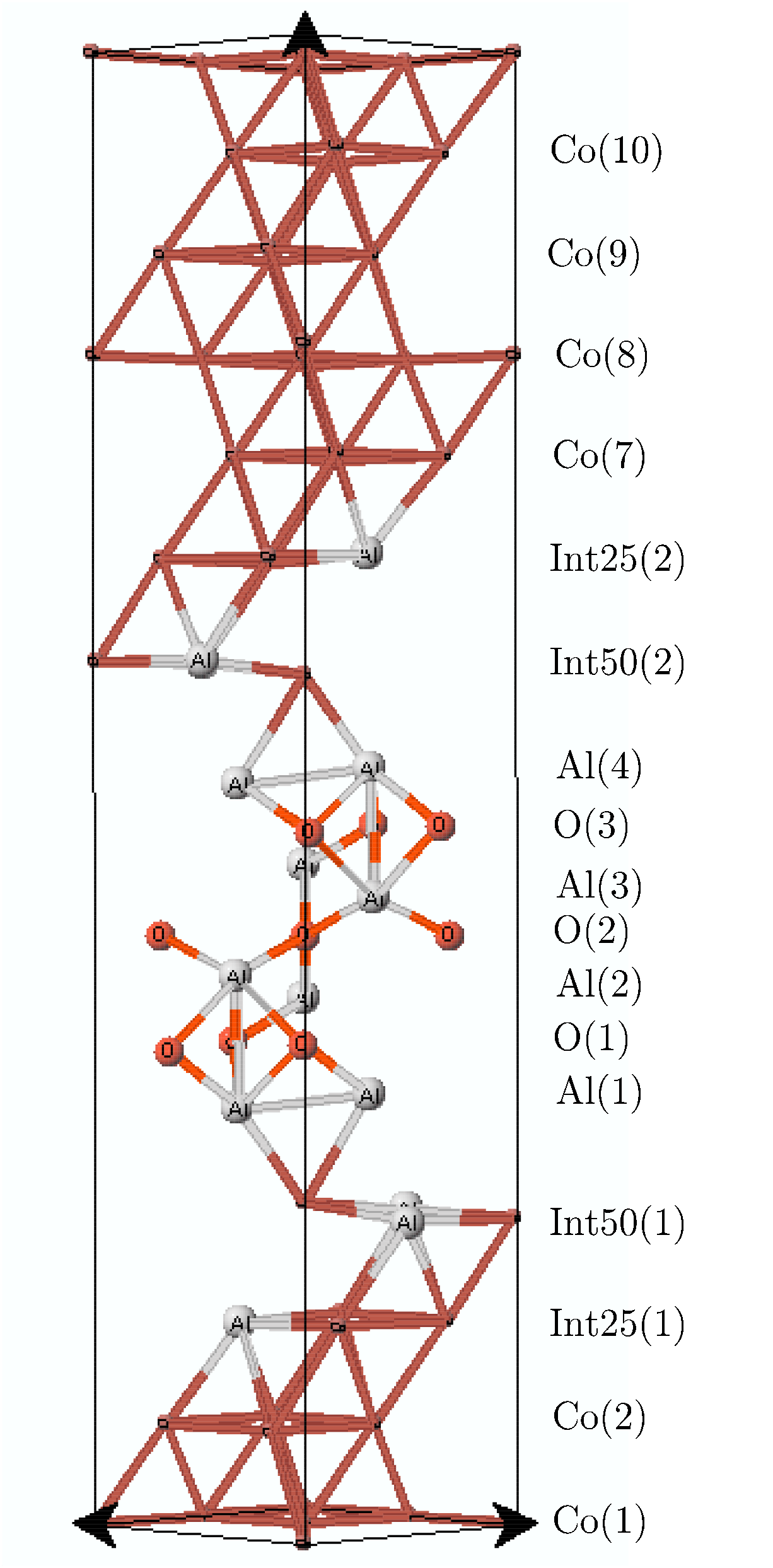}
\end{center}
\caption{Atomic structure of a disordered Al-rich interface junction. The unit vectors for the supercell as well as the layer identification tags are shown. The Int25 layers contain a 3:1 Co to Al ratio whereas the Int50 layers have a 1:1 ratio.}
\label{Supercell_2}
\end{figure}

\section{Al-Rich Disordered Interface}
\label{sec:disorderedint}
\begin{figure}[bht]
\begin{center}
\includegraphics[width=8cm]{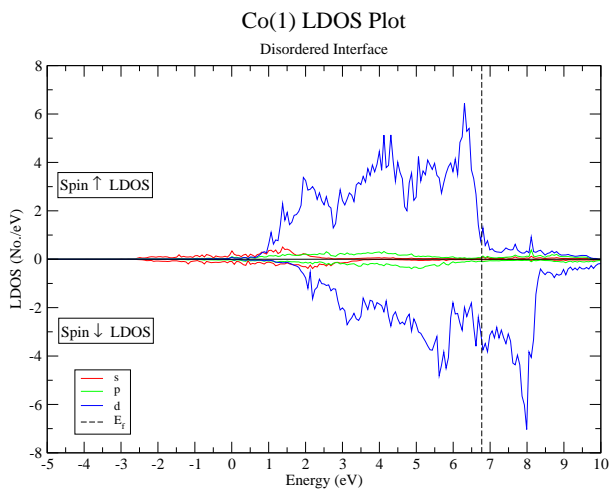}
\end{center}
\caption{Spin-dependent LDOS for Co(1) layer.}
\label{Co1_LDOS_plot}
\end{figure}

\begin{figure}[bht]
\begin{center}
\includegraphics[width=8cm]{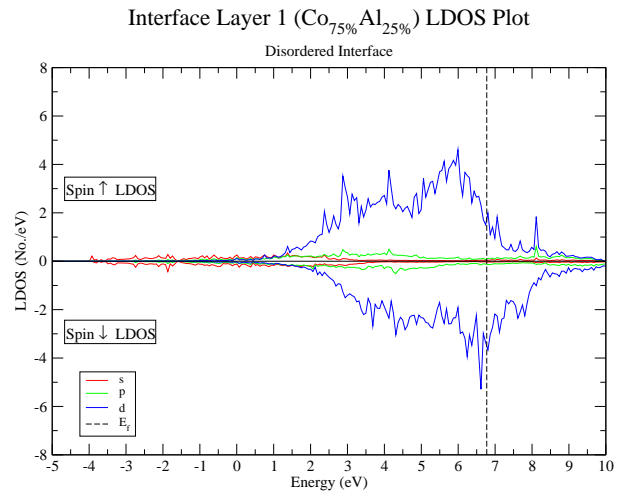}
\end{center}
\caption{Spin-dependent LDOS for first interface layer, Int25(1), with the ratio Co:Al = 3:1.}
\label{Int1_LDOS_plot}
\end{figure}

\begin{figure}[bht]
\begin{center}
\includegraphics[width=8cm]{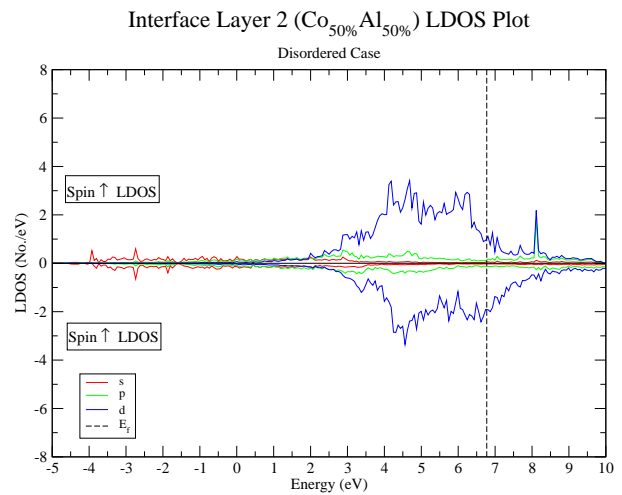}
\end{center}
\caption{Spin-dependent LDOS for second interface layer, Int50(1), with the ratio Co:Al = 1:1.}
\label{Int2_LDOS_plot}
\end{figure}

\begin{figure}[bht]
\begin{center}
\includegraphics[width=8cm]{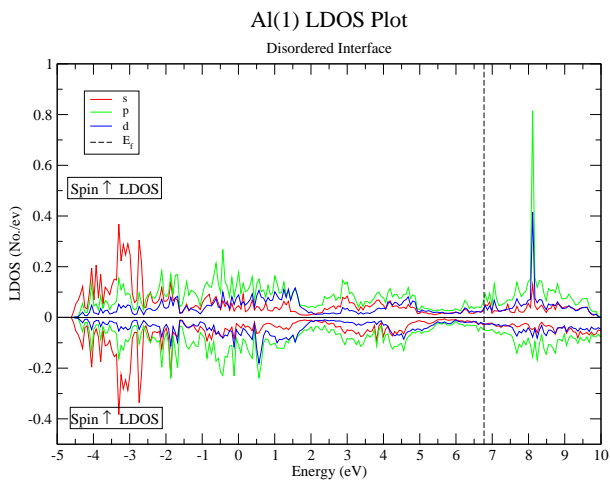}
\end{center}
\caption{Spin-dependent LDOS for first Al layer, Al(1).}
\label{Al1_LDOS_plot}
\end{figure}

The self-consistent atomic structure of the supercell for modeling the Al-rich disordered interface, with interface layers Int25 and Int50, is shown in Fig.~\ref{Supercell_2}. The first interface layer Int25(1), i.e. the layer closest to the Co substrate, has a 3:1 ratio of Co to Al, and the second mixed layer Int50(1), which is next to the Al(1) layer, has a 1:1 ratio of Co to Al. In Figs.~\ref{Co1_LDOS_plot}-\ref{Al1_LDOS_plot} we plot the spin dependent LDOS for the first Co layer, Co(1), the first interface layer, Int25(1), the second interface layer, Int(50), and the first Al layer, Al(1), respectively, to show the change in spin dependent LDOS as we approach and cross the interface. Comparing these results with those from Sec.~\ref{sec:clean}, we are able to see the effects of Friedel screening of Al atoms at the disordered interface.

As expected, the LDOS for Co(1) is almost the same as bulk Co and the Co(1) layer of the clean interface case, giving an exchange splitting of $\approx 1.5$ eV. The LDOS of Co(1) shows a large spin-down polarization at $E_F$, with $\rho_{\uparrow}(E_F) \approx 1$ eV$^{-1}$ and $\rho_{\downarrow}(E_F) \approx 4$ eV$^{-1}$, similar to that of the clean interface case. This means the screening length for the interfacial disorder effects is very short, of about one atomic layer, as expected for a metal. 


The mixing of Al and Co in the disordered layers near the interface results in spin dependent Friedel screening, which changes the LDOS. Since $\rho_{\downarrow}(E_F) > \rho_{\uparrow}(E_F)$ for bulk Co and also in the Co(1) layer, screening of the Al atoms in the Int25 and Int50 layers would be mostly from the spin-down d-electrons. Therefore $\rho_{\downarrow}(E_F)$ should change proportionally more than $\rho_{\uparrow}(E_F)$, as shown in the calculation for the Friedel screening.

The effects of the spin dependent screening is seen in both the LDOS near $E_F$ and the total number of d electrons per layer. There is a large change in the LDOS at $E_F$ for the Int50 layer compared to the Co(1) layer, which has significant importance for tunneling transport. For the Int50 layer, $\rho_{d \, \uparrow} = 1.0$~eV$^{-1}$, and $\rho_{d \, \downarrow} = 1.9$~eV$^{-1}$. Comparing this to the bulk-like layer of Co(1) where $\rho_{d \, \uparrow} = 1.0$~eV$^{-1}$ and $\rho_{d \, \downarrow} = 3.4$~eV$^{-1}$, we see that there is clearly a much larger change in the spin-down LDOS. The spin-up LDOS is unchanged, while the spin-down LDOS is almost reduced by half. Similarly, there is a much larger change for the total spin-down d electrons in the Int50 layer. The total d electron charge in the Int50 layer is 8.4~eV$^{-1}$/layer and 7.2~eV$^{-1}$/layer for spin-up and down respectively. In comparison, the average of the Al(1) and Co(1) layers is 9.0~eV$^{-1}$/layer and 6.0~eV$^{-1}$/layer respectively for spin-up and down.  This shows the spin dependent change in the LDOS of the interfacial layers is due to screening effects, whereas hybridization should lead to a similar change in the LDOS for both spins. 

In addition, we have calculated the average of the LDOS of the Co(1) and Al(1) layers in the clean case to obtain an estimate for the effects of hybridization, and compared that to the Int50 LDOS for a clearer picture of the effects of the spin-dependent screening. The LDOS near $E_F$ for the two are very similar for spin-up, whereas spin-down shows a large shift of spectral weight from around $E_F$ to the peak seen in Int50 at about 4.6 eV. This leads to the change in the spin polarization at $E_F$ and also affects the spin polarized tunneling current, both of which are discussed further in the following sections. 

The sharp peaks in the d-electron LDOS that were seen in the Co(1) layer are also significantly reduced in the Int25 and Int50 layers, indicating that the d-electron orbitals are more delocalized. This is due to hybridization with the Al s-p bands from the neighbouring Al(1) layer, and with the Al s-p electrons in the layers itself.

We notice that the first Al layer, Al(1), has a small but finite LDOS at $E_F$, since it has become metallized by proximity to the Co layers, an effect also seen for the clean interface. The effective insulating width of the Al$_2$O$_3$ layer is therefore again reduced by approximately one to two atomic layers. 

\begin{figure}[bht]
\begin{center}
\includegraphics[width=9cm]{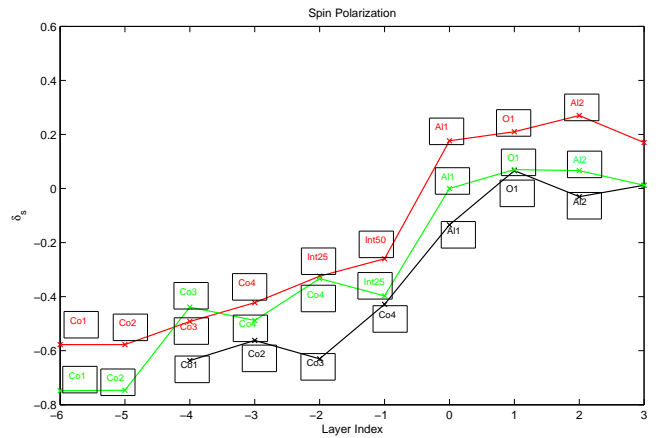}
\end{center}
\caption{Spin polarization at $E_F$, $\delta_s(E_F)$, of the three different MTJ structures investigated assigning the Al(1) layer to be layer 0. Clean interface (black), one Co/Al interfacial layer (green), and two Co/Al interfacial layers (red).}
\label{spin_polarization_plot}
\end{figure}

We plot in Fig.~\ref{spin_polarization_plot} the spin polarization, $\delta_s$, of the three different Al-terminated MTJ structure studied, starting in the bulk and moving across the interface into the Al$_2$O$_3$ insulating barrier. As we approach the interface we can see an increase in $\delta_s$ of each layer. The bulk Co layers have a $\delta_s \approx -(0.6 - 0.8)$ in all structures. For both the clean and disordered system, $\delta_s$ increases as we approach the interface. For the clean Al-terminated interface the spin polarization remains slightly negative at the interfacial Al(1) layer, and becomes slightly positive in the O(1) layer. For the two interfacial layer structure (Fig.~\ref{Supercell_2}) the spin polarization similarly increases towards the interface, $\delta_s = -0.32$ in the Int25 layer, and $\delta_s = -0.27$ in the Int50 layer. Furthermore, the spin polarization becomes positive, $\delta_s = 0.2$ , at the metallized interfacial Al(1) layer.

Looking at the change in spin polarization from the bulk-like Co(1) layer to the layer at the interface, the increase in $\delta_s$ from the Co(1) to Co(4) layer for the clean interface system can be attributed solely to hybridization with the Al s-p electron in the Al(1) layer. For this system, the spin polarization changes from -0.64 to -0.43, which is a change of 33\%. On the other hand, for the disordered system with two mixed layers, $\delta_s$ increases from -0.58 to -0.26 between the Co(1) and the Int50 layer, which is a change of 55\%. This clearly shows the additional effects of the spin dependent Friedel screening and interfacial disorder on the spin polarization. 

\begin{figure}[bht]
\begin{center}
\includegraphics[width=7cm]{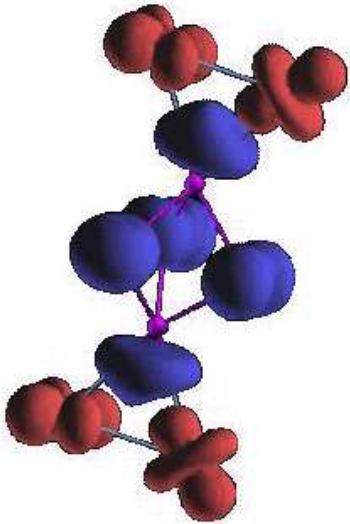}
\end{center}
\caption{Atomic structure of model system of 9 atoms representing the Al-terminated Co/Al$_2$0$_3$ MTJ structure. The figure shows the total spin polarization, and the color indicates the spin polarization for each atom. This result, matching so closely the larger calculation, is suggestive that superexchange is the mechanism for the positive spin polarization in the larger structure, and is definitely the mechanism for the observed spin polarization in this smaller structure.}
\label{small structure fig}
\end{figure}

To study the source of the positive spin polarization in the interface region shown in Fig.~\ref{spin_polarization_plot}, we have performed a calculation on a model system, an 'essential unit cell' of the interface region, shown in Fig.~\ref{small structure fig}. This is composed of an Al atom bonded to two Co atoms in one direction, and to three oxygen in the other. These oxygen are bonded to a second Al, and the structure simplified by termination at this point with two final Co. The bond lengths and angles were chosen to match those of the clean structure, Fig.~\ref{Supercell_clean}. For reasons shown in the following discussion, we believe the model structure may capture much of the essence of the spin interactions and hybridization at the interface.

Density functional calculations were performed on this system in a comparable manner to those in the rest of this paper, including full relaxation of the structure. The program Quantum Espresso \cite{QE} was used, with convergence and other properties similar to those described in Sec.~\ref{sec:compdetails}. Relaxation of the structure did not change the bond lengths or angles from those of the full calculation of Fig.~\ref{Supercell_clean}.
     
Fig.~\ref{small structure fig} shows the spatial distribution of the spin polarization, with contrasting color indicating opposite direction of spins. We note first of all the marked similarities between these results and those of Fig.~\ref{spin_polarization_plot}. The total spin polarization on both the Al and O layers is positive, opposite to the negative of the Co. In addition, the spin polarization of the Al is not centered on the Al itself, but rather is midway to the Co, indicating the extent of the hybridization. Both the Al and Co polarizations have a marked anisotropy, indicating the preferential involvement of certain orbitals in the development of the spin polarization. Al draws electrons from the Co, a process akin to Friedel screening in a larger system. The electrons near the Al are spin polarized in the opposite direction to Co due to superexchange in this small system. Similarly we find in the structure in Fig.~\ref{Supercell_clean} that the Al has a total spin polarization opposite to the Co atom. We suggest this is due to similar Pauli exclusion effects in the screening process. If the MTJ structure is O-terminated instead, the interfacial O atoms would behave similarly to the interfacial Al-atoms that we have described above; i.e. the O atom would be screened by the Co electrons and would have a net positive total spin polarization.

The spin polarization results for the three different Al-terminated systems shown in Fig.~\ref{spin_polarization_plot} shows explicitly that not only a O-rich but also a Al-rich disordered interface can give rise to a positive spin polarization in the insulating alumina layer. Moreover, this positive spin polarization can be reached in a region with a finite LDOS at the Fermi level for an Al-rich interface, namely the Al(1) layer, which is of crucial importance when studying tunneling processes. The effects of the change in spin polarization on the tunneling current is modeled using spin dependent effective masses for the different bands in each layer as discussed the section below. This is shown in Table.~\ref{table:eff mass values}, especially for the d2 bands, which contribute the most to the LDOS at $E_F$. The details of the effective mass derivations, and effects on the tunneling current are discussed in the following Sec.~\ref{sec:spincurrent}.

\section{Spin Polarized Tunneling Current}
\label{sec:spincurrent}
In the section above we showed that Al-terminated and Al-rich interfaces can give rise to a net positive spin polarization in the insulating barrier. We now explicitly show that the tunneling current is also positively spin polarized. In order to calculate the tunneling current, we model the MTJ system by a position-dependent effective mass model, which we then solve exactly to obtain the tunneling current. The tunneling layer is modeled in the standard manner as a potential barrier, $V_o$, with thickness $a$, and applied voltage $V_1$. The Hamiltonian for the system is written simply as an effective kinetic mass term and a potential barrier:

\be
H = \frac{p^2}{2 m(x)} + \theta(x)\theta(a-x)(V_0 - \frac{V_1}{a}x).
\label{eff mass Ham}
\ee

The effective mass, $m(x)$, changes discontinuously from layer to layer, and the values for the different s-p and d bands are obtained from the DFT calculations as described below. The Fermi energy is 6.78 eV, as determined from the LDA results, the effective barrier height, $V_0 = 9.4$ eV, is obtained from electron spectroscopy of thin film Al$_2$O$_3$ layers.\cite{CostinaAPL2001} The effective width of the Al$_2$O$_3$ layer, $a$, is taken to be one atomic layer thinner than the actual thickness shown in Fig.~\ref{Supercell_2}, due to the partial metallization of the Al layers at the interface. The appropriate boundary conditions to be imposed between every layer is,

\ba
\psi(0-) & = & \psi(0+) \nonumber \\
\frac{1}{m_1} \frac{d\psi}{dx}(0-) & = & \frac{1}{m_2} \frac{d\psi}{dx}(0+),
\label{boundary conds}
\ea

where $m_1$ is the effective mass for the layer on the left, $x < 0$, and $m_2$ is the effective mass for the layer on the right, $x > 0$, where $x=0$ is where the layers meet.

We extract effective masses for our model from the DFT results as follows. The DFT LDOS data is projected into s-p and d bands for each of the layers, which we model separately with different effective mass values. The d-electron band is modeled using three bands, d1, d2, and d3, so as to more realistically capture the LDOS near $E_F$. First, we project the DFT LDOS data into s-p and d bands for each of the layers and in particular look at the data near $E_F$. Next, we model the d-electron band as three separate bands in order to more accurately capture the LDOS near $E_F$. The large peak near $E_F$ is modeled using one band, labelled as d3, the overall large-effective mass d band is modeled using another band, labelled as d2, and a more flat sp-like band is labelled as d1. We model the s-p, d1, d2, and d3 bands with a free-electron density-of-states expression, with effective mass, $m^*$, and the band edge as fitting parameters.

To check our results, we also use the full LDA band structure to extract the Fermi velocity, $v_F$, for each band and then use $v_F = \tfrac{k_F}{m^*}$ to extract an effective mass, $m^*$ for each of the bands. We find that the effective masses, $m$, extracted from the free electron picture above are in good agreement with the band structure masses, $m^*$, and thus we have a high confidence in our approach. The effective masses, with respect to the bare electron mass $m_0$, are shown in Table~\ref{table:eff mass values}. Here the two disordered layers at the interface, Int25 and Int50, are treated as one single layer for ease of calculation. The LDOS of the two layers are similar, and the effective mass value used is the average of the effective masses obtained from the two layers. The d3 bands in the interface layers do not cross $E_F$ and are therefore assumed to be irrelevant for the transport calculations.

\begin{table}[!htbp]
\begin{tabular}[c]{|c|c|c|c|c|c|c|c|c|}
\hline
 & \multicolumn{4}{c}{Co(1) Layer} & \multicolumn{3}{|c|}{Int25/Int50 Layer} & Al$_2$O$_3$ Layer \\ \hline
Bands & s-p & d1 & d2 & d3 & s-p & d1 & d2 & s-p \\ \hline
 Spin-$\uparrow$   & 0.45 & 0.81 & 2.05 & 6.19 & 0.51 & 0.81 & 1.70 & 1.0 \\ \hline
 Spin-$\downarrow$ & 0.37 & 1.02 & 4.99 & 7.89 & 0.51 & 0.58 & 1.70 & 1.0 \\ \hline   
\end{tabular}
\caption{Effective mass values ($m/m_0$) for the different electron bands in the cobalt layer, the interface layer, and the insulating Al$_2$O$_3$ layer. Notice the large change in the effective mass for the spin-down d2 bands between the Co(1) and the Int25/Int50 layers.}
\label{table:eff mass values}
\end{table}

The increase in the spin polarization due to the Al-rich interface is reflected in the spin dependent effective masses of the d-electron bands, especially in the d2 bands which has the largest LDOS at $E_F$. It is shown in Table~\ref{table:eff mass values} that the d2 bands in the Co(1) layer have effective masses 2.05 $m_0$ and 4.99 $m_0$ for spin-up and down respectively, which both reduce to 1.70 $m_0$ in the interface layer. Clearly the larger decrease in spin-down LDOS near $E_F$, due to the Friedel screening effect and LDA results shown in Secs. \ref{sec:intscattering} and \ref{sec:disorderedint}, leads to a corresponding increase in spin polarization $\delta_s(E_F)$. This decrease in LDOS is modeled by the change of the effective mass values from the Co(1) layer to the interface layers. A smaller tunneling probability for the spin-down d bands hence results, due to the larger spin-down effective masses, and the greater mis-match in effective masses between the Co(1) and the Int25/Int50 layer. This is clearly seen for all three d1, d2 and d3 bands, and hence there is a smaller spin-down polarized tunneling current.

The exact wavefunctions for the Hamiltonian, Eq.~(\ref{eff mass Ham}), are described by Airy functions whose coefficients are determined numerically using the boundary conditions, Eqs.~(\ref{boundary conds}), and a normalization condition for the wavefunction. The numerically determined wavefunction is then used to calculate the tunneling probabilities, and tunneling current which is given by the standard current expression,

\begin{equation}
j = i \hbar \left(\frac{\partial \psi(x)}{\partial x} \psi^{*}(x) - \frac{\partial \psi^*(x)}{\partial x} \psi(x) \right).
\end{equation}

\begin{figure}[bht]
\begin{center}
\includegraphics[angle=-90,width=8cm]{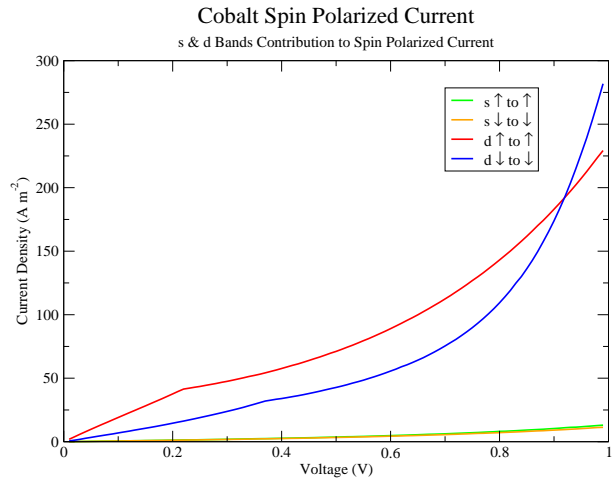}
\end{center}
\caption{Spin polarized tunneling current of MTJ structure. The lower transmission coefficient for the spin-down bands, due to a greater mis-match in effective masses, result in a lower tunneling current compared to the spin-up bands.}
\label{spin_pol_current_plot}
\end{figure}

Figure \ref{spin_pol_current_plot} shows the tunneling current for the parallel configuration, where the magnetization of the two cobalt layers are parallel. Therefore, the spin-up electrons tunnel to the spin-up bands, and similarly for the spin-down electrons. The s-p band electrons naturally tunnel to the s-p band by symmetry, and since the d1, d2 and d3 band electrons were originally modeled from the same d-electron band, they tunnel into the other bands as well; i.e. the d1 band electrons tunnel into the d2 and d3 bands as well as the d1 band. 

The spin polarized tunneling for both the s-p electrons and the d-electrons are positive, up to $\approx 0.9$ eV. The positive spin polarization tunneling current arises from a greater mis-match in the effective masses of the spin-down bands between the bulk cobalt layer, Co(1), and the interface layers as compared to a smaller mis-match in the spin-up bands. This is also seen in the s-p bands, where there is a slightly better match between the spin-up effective masses. This results in a slightly larger spin-up tunneling current for the spin-up s-p band, as shown in Fig.~\ref{spin_pol_current_plot}. There is also a large peak, modeled as the d3 band, in the LDOS of spin-up d-electrons, approximately 100 meV below $E_F$, that affects the spin polarized tunneling; whereas a similar peak for the spin-down d-electrons is 1.26 eV above $E_F$ due to the exchange splitting. This is seen clearly in Fig.~\ref{Co1_LDOS_plot}. Therefore, for an applied voltage larger than 100 meV, this large spin-up d3 band will contribute, and results in a larger spin-up polarized tunneling current. This continues until $0.91$ eV, when the bottom of the spin-down d3 band crosses the quasi-Fermi level and starts to contribute to tunneling. The positive spin polarized tunneling current reflects the positive spin polarization shown in Fig.~\ref{spin_polarization_plot} as we cross from the cobalt layers into the Al$_2$O$_3$ layer, and explains the positive spin polarization found experimentally. 

\begin{figure}[bht]
\begin{center}
\includegraphics[angle=0,width=8cm]{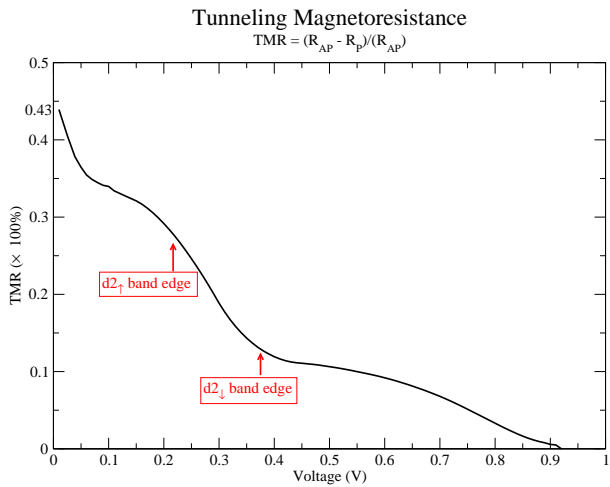}
\end{center}
\caption{Tunneling magnetoresistance of MTJ structure calculated from Julliere's formula, with spin polarization obtained from spin polarized tunneling current shown in Fig.~\ref{spin_pol_current_plot}.}
\label{plot:TMR Julliere}
\end{figure}

Finally, we can numerically calculate the tunneling spin polarization, $P = \frac{I_{\uparrow} - I_{\downarrow}}{I_{\uparrow} + I_{\downarrow}}$, where $I_{\uparrow, \downarrow}$ is the spin polarized tunneling current given in Fig.~\ref{spin_pol_current_plot}. Using Julliere's formula, Eq.~(\ref{eq:TMR Julliere}), we then calculated the TMR from the numerically derived tunneling spin polarization $P$, and the results are shown in Fig.~\ref{plot:TMR Julliere}.

The calculated spin polarization is $P = 53\%$, which compares very well with the experimental values of 42\% for Co and 55\% for Co$_{50}$Fe$_{50}$. \cite{Meservey94, Veerdonk97} The calculated TMR is 43\% at zero applied voltage, which compares favourably with the experimental value of 30\%. Next, we see that the TMR decreases monotonically with applied voltage and reaches zero at 0.91 V. The spin polarization and TMR both decrease to zero at this voltage, which is the spin-down d3 band edge and is of the order of the exchange splitting. Finally, we also see two well-defined ``roll-overs" in the TMR curve labeled ``$d2_{\uparrow}$ band edge" and ``$d2_{\downarrow}$ band edge". This is due to band structure effects; the first ``roll-over" at 0.23 V is due to the spin-up d2 band edge and the second ``roll-over" at 0.375 V is due to the spin-down d2 band edge, where the two bands stop contributing to the tunneling transport. The monotonic decrease of TMR with increase in voltage and the ``roll-overs" have been seen experimentally, see for e.g. Refs.~\onlinecite{Moodera99, MooderaPRL98, NakashioIEEE2000}. These two ``roll-overs" are much more prominent in the numerical results because the band edges for the $d2_{\uparrow}$ and $d2_{\downarrow}$ bands in the effective-mass model are modeled using step-functions, which is more abrupt than the real band structure.
 
\section{Conclusions}
In this paper, we have studied the effect of interface mixing and atomic-level disorder on the spin polarized tunneling current in Co/Al$_2$O$_3$ MTJs. The spin polarization of tunneling currents in MTJs have clearly been seen in experiments to be positive, at odds with simplistic tunneling models which depends on the spin polarization of the DOS at the Fermi energy of the leads. There is now a growing body of evidence that interface effects at the boundary between the Co layers and the aluminum oxide insulating layer play an important role in determining the spin polarized transport in these systems. A large number of studies have assumed an O-rich interface, and found positive spin polarization for such systems. However, we believe that an Al-rich/terminated interface is more relevant experimentally when the junction is not heavily oxidized.

The MTJ was modeled realistically using DFT in both the LDA and GGA approximations, of which the LDA scheme was found to be more consistently reliable. We have studied both clean Al-terminated and Al-rich disordered interface systems, and found a consistent trend of increasing positive spin polarization with Al content of the interface. The main effects of the Al-rich disordered interface are spin dependent Freidel screening of the Al-impurities in the interface lead layers, and hybridization with the s-p electrons of the Al-impurities and the neighbouring first Al layer of the insulating barrier. This causes a much larger change in the spin-down d-electron LDOS near $E_F$ of the interface layers, leading to an increase in the spin polarization near $E_F$. Another important result of our LDA calculations is that the proximity effect leads to metallization of the first Al-layer by the neighbouring Co layer. This reduces the effective width of the insulating layer, and can be seen in the slope of the I-V curve. 

Studying the interface with a model system which focuses on the Co-Al-O interactions, we find good agreement with the larger calculation on the incidence and sign of spin polarization in the interface layers. This model calculation suggests that the source of the opposite polarization of the Al and O is a superexchange mechanism similar to that seen with magnetic atoms on an insulating surface.~\cite{HeinrichScience06} Just as the RKKY interaction was originally studied with isolated magnetic impurities in bulk metals, and then widely seen in metallic magnetic multilayers, we postulate that the superexchange interactions previously seen for isolated magnetic atoms separated by a nonmagnetic insulating 'island' are also applicable to many magnetic tunnel junctions. Because the hopping interaction dies down with distance into the insulator, the net effect will be an opposite spin polarization of the interface layers.

The effects of increasing spin polarization in the LDOS near $E_F$ is captured using a layer and spin dependent effective mass model, which is used to calculate the tunneling current. The greater mis-match of the effective masses leads to a smaller transmission coefficient for the spin-down d bands, and the tunneling current clearly shows a positive spin polarization that is in agreement with experimental results. This defines an effective tunneling spin polarization, which allowed us to calculate the TMR using the Julliere model, giving results that are also in good agreement with experiments. 

In summary, using DFT calculations, we have found that Al-terminated interfaces in MTJs show positive spin polarization of the LDOS at $E_F$, showing that positive spin polarization is possible not only in O-terminated MTJ systems. Furthermore, we formulated a spin dependent effective mass model, using values extracted from the DFT results, that was used to calculate the tunneling current and the TMR. The results clearly show a positive spin polarization and TMR values that are in agreement with experiments. Thus, our model demonstrates how positive spin polarization and realistic TMR values can be obtained in an Al-terminated MTJ.

\acknowledgments{
We thank Shu Peng and Kyeongjae Cho for invaluable help in the beginning of this work. We would also like to thank Stuart Parkin and his group members for very helpful discussions and unpublished data. T. Tzen Ong was partly funded by IBM and partly by the CPN Stanford. W. Shen was funded by NSF grant DMR-0705266.}

\bibliography{MTJ_Bib_long}
\end{document}